\documentclass[12pt,letterpaper]{article}
\usepackage{amsmath,cite,amssymb,color,mathrsfs}

\newcommand{\beq}{\begin{equation}}
\newcommand{\eeq}{\end{equation}}
\newcommand{\beqa}{\begin{eqnarray}}
\newcommand{\eeqa}{\end{eqnarray}}
\newcommand{\cc}{\rm{c.c.}}
\newcommand{\cvd}{\mathscr{D}}
\newcommand{\lsim}{\mathrel{\rlap{\lower4pt\hbox{\hskip1pt$\sim$}}
    \raise1pt\hbox{$<$}}}         
\newcommand{\gsim}{\mathrel{\rlap{\lower4pt\hbox{\hskip1pt$\sim$}}
    \raise1pt\hbox{$>$}}}         

\begin{document}
\vspace{25pt}

\begin{center}

{\Large\bf  Lorentz Violation and Superpartner Masses}

\vspace{15pt}

Andrey Katz and Yael Shadmi

\vspace{7pt}
{\small
{\it{Physics Department, Technion---Israel Institute of Technology,
Haifa 32000, Israel \\
andrey,yshadmi@physics.technion.ac.il} }
\vspace{4pt} }
\end{center}

\begin{abstract}
\noindent
We consider Lorentz violation in supersymmetric extensions of the standard 
model. We perform a spurion analysis to show that, in the simplest 
natural constructions, the resulting supersymmetry-breaking masses are tiny.
In the process, we argue that one of the strongest bounds on Lorentz
violation in the photon Lagrangian, which comes from the absence
of birefringence from distant astrophysical sources, does not apply 
when Lorentz violation is parametrized by a single vector.
\end{abstract}

\section{Introduction}
Of the various ideas for physics beyond the standard model, supersymmetry
is especially appealing because it is the unique extension
of the Poincare group in four dimensions.
Given the central role of Poincare symmetry in the standard model,
and the wealth of relevant experimental data at a wide range of energies,
it is hard to imagine that it can be broken.
But precisely because of its central role, it is important
to test Poincare invariance and to understand the consequences
of its possible
breaking~\cite{Colladay:1998fq,Coleman:1998ti,AHamed:2003uy,
Cohen:2006ky}.
In this letter, we will explore the implications of this
breaking for supersymmetry breaking. 
Consider for example a chiral superfield with a scalar field $s$,
a fermion $\psi$, and an auxiliary component $F$.
Under a supersymmetry variation, the fermion transforms as
\beq \label{var}
 \delta_\xi \psi \sim \sigma^\mu \bar \xi
\partial_\mu s +\xi F ,
\eeq
where $\xi$ is the parameter of the variation.
If the Lorentz-scalar $F$ is non-zero, supersymmetry is broken with
Poincare symmetry intact. Virtually all studies of supersymmetry
breaking rely on such scalar $F$ term, or $D$-term, VEVs.
Here he will consider instead the possibility that 
$\partial_\mu s$ is non-zero, so that supersymmetry is 
broken together with Poincare 
invariance\footnote{Additional examples of the relation between Poincare 
breaking and supersymmetry breaking are discussed
in Appendix~A.}.
Can this breaking generate weak-scale soft masses?
In the ``gauged ghost condensation'' models of~\cite{Cheng:2006us},
the Lorentz breaking scale can be as high as 
10$^{15}$GeV~\cite{Cheng:2006us,Grossman:2005ej},
so it is intriguing to ask whether it can lead to soft masses that
are sufficiently large.

To answer this question, we will use a spurion analysis in
a supersymmetric extension of the standard model.
Namely, we will assume that Poincare invariance is sponatneously broken
in some hidden sector, with some field obtaining a Poincare
violating VEV.
We will then treat this field as a spurion, and 
analyze the minimal supersymmetric standard model (MSSM) 
Lagrangian in the presence of this spurion,
in order to estimate the size of the resulting soft masses
relative to the scale of Lorentz violation\footnote{With a slight abuse
of notation, we will use the term ``Lorentz violation'' in the following 
since this is the term commonly used in the literature. We note however
that in the supersymmetric setting we will consider, translations
symmetry is almost always broken by the background.}.
We emphasize that we do not consider here any additional source 
of supersymmetry breaking apart from the Lorentz violating spurion. 
It is well known of course how to generate soft masses for the MSSM 
by non-zero $F$-terms or $D$-terms.
Here we are interested in whether the dominant source
of soft masses in the MSSM can be associated with Lorentz violation,
so we assume that no other supersymmetry-breaking spurions,
namely, nonzero $F$-terms and $D$-terms, exist.

In section~\ref{chiralspurion}, we will consider a spurion which
resides in a chiral superfield as in eqn.~(\ref{var}).
Terms involving only regular derivatives of the spurion superfield
will reproduce the results of~\cite{GNibbelink:2004za},
which imposed only supersymmetry on the MSSM fields, and studied
the Lorentz structure of the Lagrangian\footnote{Because of the spurion
we use, our analysis will only reproduce those terms 
of~\cite{GNibbelink:2004za} that can be generated with 
a single Lorentz violating vector.}.
However, terms with superspace derivatives acting on the spurion
will generate soft terms, as well as explicit supersymmetry breaking
terms for the MSSM fields.
As we will show, the suppression of scalar masses is identical
to the suppression of Lorentz violation in the fermion kinetic terms.
Bounds on these Lorentz violating terms then imply that the soft masses 
generated are tiny.
Interestingly, all other Lorentz violating operators are consistent
with weak scale (or higher) soft masses.

In the simple model we will construct, CPT is automatically
conserved. It is interesting to note that supersymmetry breaking from
Lorentz violation   
is similar to $D$-term breaking, in that it preserves $R$ symmetry.
As mentioned above, the soft masses will be constrained
by Lorentz violation in the fermion sector.
Lorentz violating operators in the photon sector are not 
problematic. As we will show in section~\ref{photon}, 
the most stringent bound on these operators, 
which comes from birefringence, does not apply for models in which 
Lorentz violation can be parametrized by a single vector field vacuum
expectation 
value (VEV), as is the case here.

In section~\ref{phase} we will discuss the VEV of a
Goldstone boson field~\cite{Graesser:2005ar}.
Such constructions seem promising candidates for supersymmetric
generalizations of gauged ghost condensation models~\cite{inprogress}.

For completeness we list the relevant spurion couplings to the MSSM fields
in  Appendix~B.


\section{Chiral superfield spurion}\label{chiralspurion}
We first consider Lorentz violation from a coordinate-dependent 
scalar field VEV. If this scalar field is the lowest component of a
chiral superfield $S$ then,
 \beq
 S=s(x^\mu)-i\theta\sigma^\mu \bar \theta 
\partial_\mu s~ +\cdots
\eeq
where the dots stand for terms involving the fermion and auxiliary
components. 
We take 
 \beq
 \langle \partial_\mu s \rangle =e^{i\phi} E^2 N_\mu ,
 \eeq
where $N_\mu$ is a timelike unit vector, $E$ is some energy scale,
and $\phi$ is a constant phase. 
There is then a special ``ether'' frame in which 
$N_\mu=(1,0,0,0)$, and the rotation group is unbroken.

Since the $s$ VEV is coordinate dependent, 
direct couplings of $S$ to the MSSM fields would lead to
explicit coordinate dependence. 
To avoid this we impose a shift symmetry 
following~\cite{AHamed:2003uy}, 
 \beq \label{supershift}
 S\to S+A\ 
 \eeq
where $A$ is a constant.
The Lagrangian therefore contains only regular derivatives and superspace
derivatives of $S$, which give two Lorentz breaking spurions,
\beqa\label{spurions}
\langle \partial_\mu S\rangle &=& e^{i\phi} E^2 N_\mu\nonumber\\
\langle D_\alpha S\rangle &=& -2 i e^{i\phi} \sigma^\mu_{\alpha\dot\alpha}
\bar\theta^{\dot\alpha} E^2 N_\mu\ .
\eeqa
The VEVs of higher derivatives of $S$ either vanish, or can be written in terms
of~(\ref{spurions}). 
In the following, we will examine the possible couplings
of $S$ to the MSSM fields. 
These will give rise to soft supersymmetry-breaking terms, as well as
to Lorentz violating terms.
Note that, as in $D$-term breaking, supersymmetry can be broken without
$R$ symmetry breaking, since we can choose $S$ to have zero $R$-charge.
Therefore, gaugino masses are either zero or suppressed compared to
squark masses.

\subsection{Superpotential}
Since the field $DS$ is not chiral it can not appear in the
superpotential.  
The superpotential therefore involves only regular derivatives of $S$
\beq
 W=W(\partial_\mu S)~.
 \eeq

Consider the MSSM chiral field $Q$. Its regular derivative
$\partial_\mu Q$ is chiral, but not gauge invariant. 
The covariant derivative,
\beq
{\cal D}_\mu\equiv -\frac{i}{4}\bar
{\sigma_\mu}^{\dot \alpha \alpha }\bar D_{\dot \alpha}{\cal
D}_{\alpha}
\eeq
 with
 \beq
 {\cal D}_\alpha Q\equiv e^{-V}D_\alpha (e^V Q)~,
\eeq
is not chiral however, and therefore cannot
appear in the superpotential. This means that we can not
contract $\partial_\mu S$ with derivatives of the MSSM superfields
in the superpotential.

Couplings of the form
\beq\label{ww}
W \sim \partial^\mu S \partial_\mu  W_0~,
\eeq
where $W_0$ is some gauge invariant combination of the MSSM fields, 
are allowed, but would give total derivative terms.
The only relevant couplings therefore involve self-contractions of 
$\partial_\mu S$,
\beq\label{nrs}
W\sim (\partial_\mu S \partial^\mu S)^{n}\, W_0~,
\eeq
where $W_0$ depends on the MSSM superfields.
Such terms merely give Lorentz-preserving renormalizations of the 
MSSM superpotential.
Therefore, the superpotential does not give rise to either soft terms
or to Lorentz violation.
In the following we will impose an additional global $U(1)$ symmetry
on $S$ in order to forbid dangerous Lorentz violating terms in 
the K\"ahler potential.
Since $S^\dagger$ cannot appear in the superpotential, 
the couplings~(\ref{ww}) and~(\ref{nrs}) are then forbidden,
and the superpotential does not contain any $S$ couplings to the MSSM.

\subsection{K\"ahler potential: the minimal model}
The K\"ahler potential is a lot less restricted. 
It is easy to see that the lowest order term that leads to 
scalar masses is
\beq
K\sim \frac{1}{M^6}D^\alpha S D_\alpha S \bar D_{\dot \alpha }
S^\dagger \bar D^{\dot \alpha} S^\dagger Q^\dagger Q\ ,
\eeq
where $Q$ denotes an MSSM matter superfield, and $M$ is the high scale
at which Lorentz violation is communicated to the MSSM. We will
usually take this 
scale to be the Planck scale.
The resulting scalar masses are
\beq\label{squarkmass}
\tilde m^2_{\tilde q}\sim \left(\frac EM \right)^8 M^2\ .
\eeq
Gaugino masses require explicit $R$-symmetry breaking,
and are generated at lowest order by,
\beq
K\sim \frac{1}{M^7}  D^\alpha S D_\alpha S \bar D_{\dot \alpha }
S^\dagger \bar D^{\dot \alpha} S^\dagger\, W^\alpha W_\alpha +\cc ,
\eeq
giving
\beq\label{gauginomass}
m_{\tilde g}\sim \left(\frac EM \right)^8 M~.
\eeq

The full list of $S$ couplings to the MSSM (to lowest order) appears
in Appendix~B.
We can forbid many dangerous Lorentz-violating terms in this list
by imposing a global $U(1)$ symmetry under which $S$ is charged, with all MSSM
fields neutral.
The only surviving terms then contain pairs of $S$ and $S^\dagger$,
so that the vector $N_\mu$ only appears an even number of times.
As a result, CPT is preserved. 
In fact, if we also require $R$-invariance, the only other surviving terms are 
\beqa \label{kinren}
 \frac{1}{M^4} \partial_\mu S \bar D_{\dot \alpha}
  S^\dagger Q^\dagger e^V \bar \sigma^{\mu \dot \alpha \alpha }{\cal
  D}_{\alpha} Q~,
  \\  \label{kinmod}
\frac{1}{M^4} D_\alpha S \bar D_{\dot \alpha} S^\dagger
  \bar {\cal D}^{\dot \alpha} Q^\dagger e^V {\cal D}^\alpha Q~.
\eeqa
For any {\em Abelian} gauge field we can also write down the following
R-invariant operator:
\beq\label{photmod}
\frac{1}{M^4} (D^\alpha S W_\alpha) (\bar D^{\dot
  \alpha}S^\dagger \bar W_{\dot \alpha})~.
\eeq

Despite its strange form, the operator~(\ref{kinren}) does not
mediate Lorentz violation to the visible sector.  
Its explicit contribution to the Lagrangian is: 
\beq
{\cal L}\sim 4\left(\frac EM\right)^4 \bar \psi\sigma^\mu \cvd_\mu
\psi-4i \left(\frac EM\right)^4(\cvd_\mu q^* \cvd^\mu
q+|F|^2)~,
\eeq
where $\cvd$ denotes the usual covariant derivative with respect to the 
standard model gauge group.
Thus, this operator just gives harmless renormalizations of 
the kinetic terms for both bosons and fermions.

The operator~(\ref{kinmod}) mediates both Lorentz and supersymmetry
breaking to the visible sector. 
It gives rise to the Lagrangian
\beqa\label{cij}
{\cal L}=&-8 i C \left(\frac E M\right)^4 \psi \sigma^\mu\cvd_\mu \bar
\psi -8iC \left(\frac{E}{M}\right)^4N_\mu N_\nu \psi\sigma^\mu
\cvd^\nu \bar \psi -\\\nonumber
&16 C \left(\frac{E}{M}\right)^4N_\mu N_\nu
\cvd^\mu \tilde {q^*} \cvd^\nu \tilde q~,
\eeqa
where we introduced the numerical coefficient $C$, which is
actually a matrix in flavor space.
This term manifestly violates supersymmetry. 
While its contribution to the fermion kinetic term contains both Lorentz
violating and Lorentz preserving terms, its contribution to the
scalar kinetic term is completely Lorentz violating.

The suppression of these terms is equal to the
suppression of squark masses (see eqn.~(\ref{squarkmass})).
The numerical coefficients $C$ would be different generically for 
left-handed and right-handed fields of the standard model.  
We therefore denote them by $C_L$ and $C_R$.
In terms of 4d Dirac spinors, using the notations of~\cite{Colladay:1998fq},
the second term of eqn.~(\ref{cij}) then leads to the following
Lorentz-violating  modification of the fermion kinetic terms,
\beq
{\cal L}=ic_{\mu \nu} \bar \Psi \gamma^\mu \cvd^\nu \Psi+id_{\mu
  \nu} \bar \Psi \gamma^\mu \gamma^5 \cvd^\nu \Psi
\eeq
with
\beqa
c_{\mu \nu}&=& 4\left(\frac EM\right)^4N_\mu N_\nu (C_L+C_R)\\
d_{\mu \nu}&=& 4\left(\frac EM\right)^4N_\mu N_\nu (C_L-C_R)~.
\eeqa
Note that the standard model gauge group structure imposes the condition
$C_L^{(u)}=C_L^{(d)}$. 
In principle, the matrices $c_{\mu \nu}$ and $d_{\mu \nu}$
have non-diagonal elements in flavor space
and cause flavor-changing Lorentz violation which is severely 
constrained~\cite{Coleman:1998ti}. 
In the following we will concentrate on the diagonal terms,
since, as we will see, even these are very problematic.

Combining these results with~(\ref{squarkmass}) and assuming no
accidental cancellation between the left-handed and
right-handed parameters we conclude that
\beqa
c_{00}&\sim d_{00}&\sim\frac{\tilde m_{\tilde q}}{M}\nonumber\\ 
c_{0J}&\sim d_{0J}&\sim\beta \frac{\tilde m_{\tilde q}}{M}\\  
c_{JK}&\sim d_{JK}&\sim \beta^2 \frac{\tilde m_{\tilde q}}{M}~.\nonumber 
\eeqa
Here $\beta$ is the velocity of the earth (where the relevant 
measurements are performed) relative to the ether frame,
and J,K are the spatial directions.
The clock comparison experiment~\cite{Kostelecky:1999mr} gives
$|c_{JK}|\leq 10^{-27}$ for the neutron.
A similar bound should hold for $c_{JK}$ at the quark level. 
Since the Earth's speed relative to the ether is presumably at least as large
as its speed relative to the CMBR, we take $\beta\sim10^{-3}$, 
so that the squark mass is bounded by
\beq\label{constr}
\tilde m_{\tilde q}\lsim 10^{-21} M~.
\eeq
Even if $M$ is the Planck scale, this is four or five orders 
of magnitude too low!

The experimental bounds on $d_{\mu \nu}$ are even stronger. 
For the electron~\cite{Bluhm:1999ev},
\beq
|\tilde b(e^-)_{X,Y}|\leq 10^{-29} {\rm GeV}~,
\eeq
which translates into (see ref.~\cite{Bluhm:1999ev} for the definition of
$b(e^-)_{X,Y}$)
\beq
|d_{0J}|\leq 10^{-29}\ \ \ \rm{and} \ \ \ \tilde m \lsim
10^{-26}M~. 
\eeq
However, this bound is not necessarily applicable in our theory. 
If the hidden and visible sectors only couple through
gravitational loops~\cite{Cheng:2006us}, 
the coefficients $C_{ij}$ of eqn.~(\ref{cij}) would be 
identical for all the MSSM fields, and 
$d_{\mu \nu }$, which is proportional to $C_L-C_R$, would vanish.

Our spurion analysis reproduces many of the operators 
of~\cite{GNibbelink:2004za}, 
which wrote down a Lagrangian for the standard model 
fields (taken to transform in the usual representations of the
full Poincare plus supersymmetry algebra)
imposing only supersymmetry and translation invariance.
Lorentz invariance then emerges as an approximate symmetry of the
low-energy theory, with Lorentz violating terms appearing
at dimension-5 and higher.
This Lagrangian is simply reproduced in our analysis by terms
involving different powers of $\partial_\mu S$ 
(we can only reproduce
of course terms that can be generated with a single vector spurion).
For example the K\"ahler potential operator 
\beq
\frac{1}{M}Q^\dagger e^V {\cal D}_\mu Q
\eeq
of~\cite{GNibbelink:2004za}
originates from
\beq
\frac{1}{M^3} \partial^\mu S Q^\dagger e^V {\cal D}_{\mu}Q~.
\eeq

\subsection{Lorentz violation in the photon Lagrangian}
\label{photon}
As we mentioned above, once we impose the global U(1) symmetry, the
lowest order Lorentz violating operator in the gauge sector is the photon 
operator~(\ref{photmod}). 
Interestingly however, the Lorentz violating terms arising from this
operator alone do not constrain significantly the scale of Lorentz
violation, and are in fact consistent with weak-scale scalar masses.
The main reason for this is that the most stringent bound on Lorentz
violation in the photon Lagrangian, from birefringence of light
from distant astrophysical sources~\cite{Kostelecky:2001mb}, simply
does not apply here, or more generally, 
in models where Lorentz violation can be parametrized by a single vector.
As shown in~\cite{Kostelecky:2001mb},
the Lorentz violating photon Lagrangian\footnote{While we cannot write
the analogues of~(\ref{photmod}) for non-Abelian gauge fields, 
Lorentz violating terms such as~(\ref{LVphoton}) will be
generated radiatively, 
from ``vacuum polarization'' amplitudes with Lorentz violation and
supersymmetry breaking insertions.}
\beq\label{LVphoton}
\Delta {\cal L}=(k_F)_{\mu\nu\rho\sigma}\, F^{\mu\nu}\, F^{\rho\sigma}
\eeq
with arbitrary $k_F$, leads to different dispersion relations
for the two independent photon polarizations
\beq
E_{\pm}=(1+\rho\pm \sigma)|\vec p|~,
\eeq
where 
\beq
\rho=-\frac12 \tilde k_{\alpha}^{\phantom{i}\alpha} \,, \ \ \ \ \ {\rm and}\ 
 \sigma=\sqrt{\frac12 (\tilde k_{\alpha \beta})^2-\rho^2}\ ,
\eeq
with
\beq
\tilde k^{\alpha \beta}=(k_F)^{\alpha \mu \beta \nu }\frac{p_\mu
  p_\nu}{|\vec p|^2}~.
\eeq
The polarization of light from a celestial object is 
then wavelength-dependent.
Since the effect is also proportional to the distance from the light source, 
refs.~\cite{Kostelecky:2001mb} use measurements of
polarized light coming from distant objects to bound some components of
$k_F$ at the level of 10$^{-32}$.

However, in our case, and whenever Lorentz violation is due to
a single vector,
\beq\label{kten}
(k_F)_{\mu\nu\rho\sigma}
\sim \left(\frac{E}{M}\right)^4
N_{[\mu}\eta_{\nu][\rho} N_{\sigma]}~.
\eeq  
It is easy to see that for $k_F$ of this form $\sigma$ vanishes 
identically\footnote{To leading order in $k_F$.}.
Therefore, whenever Lorentz violation can be parametrized by a single
vector VEV, 
it is not constrained by birefringence.

The next relevant bound on $k_F$ comes from cavity resonator 
experiments~\cite{Kostelecky:2001mb,Muller:2004zp}
which give $k_F\lsim 10^{-15}-10^{-16}$.
This gives $\beta^2 (E/M)^4 \lsim 10^{-16}$,
and therefore  $m_{\tilde q}\lsim 10^{-10} M$, which can be
very high. 
We note however that the analysis of~\cite{Muller:2004zp} assumes that
some components of the tensor $k_F$ are zero, motivated by
the birefringence bound discussed above.

\section{A Goldstone--or vector field--spurion}\label{phase}
The Lorentz violating operators we considered in the previous section
were 
suppressed by large powers of the high scale $M$, 
because they had to involve derivatives of the spurion.
Another way to implement the shift symmetry
is to consider a coordinate-dependent VEV of some Goldstone
boson~\cite{Graesser:2005ar}.
The simplest example is a chiral superfield $G$,
charged under a global U(1),
whose lowest component has a coordinate-dependent VEV of the form
\beq
\langle g\rangle =\Lambda e^{i\phi(x)}\ ,
\eeq
with
\beq
\partial_\mu \phi =E\, N_\mu~.
\eeq
The spurion chiral superfield is then
\beq
G=\Lambda e^{i\phi}+E\Lambda N_\mu \theta \sigma^\mu \bar \theta e^{i
  \phi }-\frac14 E^2 \Lambda e^{i \phi} \theta \theta \bar \theta
  \bar
\theta~.
\eeq
Because of the U(1) symmetry, only the combination $G^\dagger G$,
which is coordinate-independent, is allowed.
In fact, this choice seems like a good starting point for
supersymmetrizing 
the ``gauged ghost condensation'' construction, since we can simply gauge
the U(1) symmetry. 
The basic invariant to consider is then
\beq
G^\dagger e^U G~,
\eeq
where $U$ is the U(1) gauge superfield.
Clearly, from the point of view of our spurion analysis, 
there is no difference between a vector field spurion,
\beq\label{vct} 
U=\ldots+\theta \sigma^\mu \bar \theta u_\mu+\ldots ~, 
\eeq
with
\beq 
\langle u_\mu \rangle=E N_\mu~,
\eeq
and the combination $G^\dagger G$, so for concreteness
we will consider the latter in the following.

Since we must have both $G$ and $G^\dagger$, no Lorentz violating
operator 
can appear in the superpotential.
The lowest-dimension K\"ahler potential operator coupling the spurion to the
MSSM fields is
\beq
K\sim \frac{G G^\dagger Q^\dagger Q}{M^2}~.
\eeq
This operator gives supersymmetry-breaking scalar masses of order,
\beq\label{scalarmass2}
\tilde m_{\tilde q}\sim \frac{E\Lambda }{M^2}\,M~.
\eeq
However, it also induces Lorentz violating fermion ``mass'' terms
\beq
{\cal L}\sim \frac{\Lambda^2 E}{M^2}\psi^\alpha (N\cdot
\sigma)_{\alpha \dot \alpha} \bar \psi^{\dot \alpha}\ .
\eeq
Unlike in the previous section, this term contains a single power
of $N_\mu$ and is CPT odd. 
It is therefore severely constrained, both by 
neutral-meson oscillations~\cite{Kostelecky:1999bm}, and by
 terrestrial clock experiments~\cite{Bluhm:1999ev}.
The latter give
\beq
\frac{\Lambda^2 E}{M^2}\lsim 
10^{-29}{\rm GeV}\ .
\eeq 
For the scalar mass~(\ref{scalarmass2}) to be of order a TeV,
we then need $E$ much higher then the Planck scale.

\section{Conclusions}
In this letter, we studied the implications of Lorentz
violation in some hidden sector for supersymmetry breaking
in the MSSM.
It is easy to see that Lorentz violating VEVs can simply mimic
$F$-term or $D$-term supersymmetry breaking.
The Lorentz violating spurion typically involves $\theta \sigma^\mu
\bar\theta$, 
so its square gives the usual $\theta^2\bar\theta^2$ contribution.
This depends of course on the form of Lorentz violation.
For example, for a lightlike vector $N_\mu$, the soft masses
we found vanish, because they are proportional to $N^2$,
whereas supersymmetry breaking terms such as eqn.~(\ref{cij}) remain.
As in $D$-term breaking, Lorentz violation can lead to supersymmetry
breaking without $R$-symmetry breaking. 

Even though the combination of supersymmetry and an additional
U(1) symmetry can forbid the most dangerous Lorentz violating terms
for the standard 
model fields, and in particular, CPT-odd terms, bounds on Lorentz
violating 
fermion terms imply that the resulting contributions to scalar
and gaugino masses are tiny.
One can also imagine models in which direct couplings of the MSSM
to the Lorentz violating spurion are forbidden, and supersymmetry breaking is
transmitted to the MSSM through messenger fields as in gauge-mediation
models~\cite{Dine:1994vc}.
Lorentz violating terms for the MSSM fields would then be
generated radiatively, 
and again, the bounds on these would probably imply that
the allowed supersymmetry-breaking terms are very small.
Still, such a setup would have the advantage that all supersymmetry
breaking and Lorentz violating terms in the MSSM would be flavor
blind. As we saw, this avoids many bounds on Lorentz violation.

We did not study here the origin and dynamics of Lorentz violation.
It would be interesting to extend the analyses of~\cite{AHamed:2003uy,
Cheng:2006us} to the supersymmetric case, 
and we pointed out one possible starting point for such an
analysis.

\vskip1cm
\noindent
{\bf Note added:} After this work was completed, ref.~\cite{Cohen:2006sc}
appeared, which extends Very Special Relativity to supersymmetry,
with ``half'' the supersymmetry preserved.

\vskip1cm
\noindent
{\bf Acknowledgements}\\
We thank Nima Arkani-Hamed, Michael Dine, Yuval Grossman, Yossi Nir,
Lisa Randall, Yuri Shirman and especially
Ari Laor for useful discussions.
Research supported by the United States-Israel Science Foundation
(BSF) under grant 2002020 and by the Israel Science Foundation (ISF)
under grant 29/03. 

\appendix
\section{Different Lorentz Violating Spurions}
We are interested in sponatnous Poincare breaking in 4d theories
with ${\cal N}=1$ global supersymmetry.
The relevant Poincare-breaking VEVs can therefore appear  either in the chiral
supermultiplet, or in the vector supermultiplet.
Consider first the chiral supermultiplet. The only possible Poincare
breaking it can contain is from a nonzero VEV of $\partial_\mu s$, where
$s$ is the scalar field. The variation of the fermion is then given 
by~(\ref{var}) and is nonzero, so that supersymmetry is broken.
(This type of spurion is considered in Section~2.)

In the vector multiplet, there are a few possibilities.
First, the field strength $F_{\mu\nu}$ could obtain a VEV.
Then, the variation of the gaugino is non-zero,
since
\beq
\delta_\xi\lambda =i \xi D +\sigma^{\mu\nu}\xi F_{\mu\nu}\ ,
\eeq
where $\lambda$ is the gaugino and $D$ is the auxiliary field.

Second, the gauge boson $A_\mu$ could get a VEV. 
If this vector is associated with an unbroken gauge symmetry,
this VEV is unphysical and can be rotated away.
If however the vector is associated with a spontaneously broken 
gauge symmetry, 
then the $A_\mu$ VEV can be gauge rotated
into a coordinate-dependent scalar field VEV.
(This type of spurion is considered in Section~3.)

In principle, one could also consider a massive vector field.
Then one cannot go into the Wess-Zumino gauge,
and the fermion which is the theta component of the vector
is physical.
The supersymmetry variation of this fermion contains the term
\beq
\xi \sigma^\mu A_\mu\ ,
\eeq
which is non-vanishing if $A_\mu$ is non-zero.

Note that in all these examples, half the supersymmetry may be preserved,
depending on the choice of VEV.
For example, in eqn.~(\ref{var}), variations ``orthogonal''
to $\sigma^\mu  \partial_\mu s$ vanish, 
as appropriate for 3d supersymmetry.

\section{The list of operators}
We now list the leading K\"ahler potential terms that couple
the MSSM fields to the spurion of section~\ref{chiralspurion}.
We omit operators that merely renormalize the usual MSSM K\"ahler terms.

{\bf Single $S$, chiral MSSM superfields:}
\beqa \label{op1}
& \left({1}/{M^3}\right)& \partial_\mu S \, Q^\dagger e^V {\cal D}^\mu 
  Q +\cc  \\ \label{op2}
& \left({1}/{M^2}\right)& D_\alpha S \, Q^\dagger e^V {\cal D}^\alpha Q
  +\cc 
\eeqa

{\bf Single $S$, MSSM gauge superfields:}
\beqa \label{op3}
&\left({1}/{M^3}\right)& \partial_\mu S \, W^\alpha \sigma^\mu_{\alpha \dot
  \alpha }\bar W^{\dot \alpha} +\cc\\ \label{op4} 
&\left({1}/{M^3}\right)& D^\alpha S \, W^\beta {\cal D}_\alpha W_\beta
  +\cc \\\label{op5}
&\left({1}/{M^4}\right)& \partial^\mu S \, W^\alpha \partial_\mu
  W_\alpha +\cc \\\label{op6}
&\left({1}/{M^4}\right)&\partial^\mu S \sigma_{\mu \alpha \dot
  \alpha } \, W^\beta \bar {\cal D}^{\dot \alpha} {\cal D}_\alpha W_\beta
  +\cc
\eeqa

{\bf $S^2$, MSSM chiral superfields:} 
\beqa \label{op7} 
&\left({1}/{M^3}\right)& D^\alpha S D_\alpha S\,
  Q^\dagger e^V Q +\cc\\\label{op8}
&\left({1}/{M^4}\right)& D_\alpha S \bar D_{\dot \alpha} S^\dagger \, 
  \bar {\cal D}^{\dot \alpha} Q^\dagger e^V {\cal D}^\alpha
  Q+\cc\\\label{op9}    
&\left({1}/{M^4}\right)& \partial_\mu S \bar D_{\dot \alpha}
  S^\dagger \, Q^\dagger e^V \bar \sigma^{\mu \dot \alpha \alpha }{\cal
  D}_{\alpha} Q+ \cc\\\label{op10} 
&\left({1}/{M^5}\right)&\partial_\mu S^\dagger D_\alpha S \, {\cal
    D}_\mu 
  Q^\dagger e^V {\cal D}^\alpha Q +\cc  
\eeqa

{\bf $S^2$, MSSM gauge fields:}
\beqa \label{op11}
&\left({1}/{M^6}\right) & \partial_\mu S \partial_\nu S^\dagger \,
\partial^\mu W^\alpha 
 \sigma^\nu_{\alpha \dot \alpha} \bar W^{\dot
   \alpha}+\cc\\\label{op12} 
&\left({1}/{M^4}\right)& \partial_\mu S \partial_\nu S^\dagger \,
 W^\alpha   
  {\sigma^{\mu \nu }}_{\alpha}^{\beta} W_\beta +\cc \\\label{op13}
&\left({1}/{M^4}\right)& D^\alpha SD_\alpha S \, W^\beta W_\beta
  +\cc\\ \label{op14}
&\left({1}/{M^4}\right)& D^\alpha S W_\alpha \, \bar D^{\dot
  \alpha}S^\dagger \bar W_{\dot \alpha} +\cc\\\label{op15}
&\left({1}/{M^5}\right)& D^\beta S \partial^\mu S^\dagger \sigma_{\mu 
    \alpha \dot  
  \alpha } \, \bar W^{\dot \alpha } {\cal D}_\beta W^\alpha +\cc\\
  \label{op16} 
&\left({1}/{M^6}\right)& D^\beta S \partial^\mu S^\dagger \, W^\alpha
  \partial_\mu {\cal D}_\beta W_\alpha 
\eeqa

{\bf  $S^3$, MSSM chiral superfields:}
\beqa \label{op17}
&\left({1}/{M^5}\right)&  D^\alpha S {\sigma^\mu}_{\alpha \dot
  \alpha} \bar D^{\dot \alpha}S^\dagger \partial_\mu S \, Q^\dagger
e^V Q +\cc \\ \label{op18} 
&\left({1}/{M^6}\right)& (D^\alpha S D_\alpha S +\bar D_{\dot
    \alpha} 
  S^\dagger \bar D^{\dot \alpha} S^\dagger) \,
  \partial_\mu(S+S^\dagger)\, 
  Q^\dagger e^V {\cal D}_\mu Q \\\label{op19}
&\left({1}/{M^5}\right)& D^\alpha S D_\alpha S \bar D_{\dot \beta} 
  S^\dagger \, {\cal D}^{\dot \beta} Q^\dagger e^V Q+\cc\\\label{op20}
&\left({1}/{M^6}\right)& DS \sigma^\mu \bar D S^\dagger D_\beta
  S \, {\cal D}_\mu Q^\dagger e^V {\cal D}^\beta Q \\ \label{op21}
&\left({1}/{M^7}\right)& D^\alpha S D^{\dot \alpha} S^\dagger
  \partial_\mu S \, {\cal D}_{\dot \alpha}Q^\dagger e^V {\cal D}^\mu
  {\cal D}_\alpha Q 
\eeqa

{\bf $S^3$, MSSM gauge superfields:}
\beqa\label{op22}
&\left({1}/{M^6}\right) & DS\sigma^\mu \bar DS^\dagger\partial_\mu
S \, W^\alpha W_\alpha +\cc\\\label{op23}
&\left({1}/{M^6}\right)&D^\alpha SD_\alpha S \partial_\mu S\,
W\sigma^\mu \bar W+\cc\\\label{op24}
&\left({1}/{M^6}\right) & \bar D_{\dot \alpha}S^\dagger \bar
D^{\dot \alpha} S^\dagger D^\alpha S \, W^\beta {\cal D}_\alpha
W_\beta\\\label{op25} 
&\left({1}/{M^7}\right) & D^\alpha SD_\alpha S \partial^\mu S\,
W^\beta \partial_\mu W_\beta+\cc\\\label{op26} 
&\left({1}/{M^7}\right) & D^\alpha S D_\alpha S \bar D_{\dot \beta} 
S^\dagger \, \bar W^{\dot \beta} W^\gamma W_\gamma 
\eeqa

{\bf $S^4$:}
\beqa\label{op27}
&\left({1}/{M^6}\right) & D^\alpha S D_\alpha S \bar D_{\dot
  \alpha}S^\dagger \bar D^{\dot \alpha}S^\dagger \, W^\beta W_\beta
  +\cc\\\label{op28}
&\left({1}/{M^6}\right) & D^\alpha S D_\alpha S \bar D_{\dot
  \alpha}S^\dagger \bar D^{\dot \alpha}S^\dagger \,Q^\dagger e^V
  Q\\\label{op29}  
&\left({1}/{M^8}\right)& \partial_\mu S D_\alpha S \bar D_{\dot
  \alpha }S^\dagger \bar D^{\dot \alpha}S^\dagger \, {\cal D}^\mu
  Q^\dagger e^V {\cal D}^\alpha Q+\cc\\\label{op30}
&\left({1}/{M^8}\right)& \bar D_{\dot \alpha} S^\dagger \bar
  D^{\dot \alpha} S^\dagger D^\alpha S \partial^\mu S
  {\sigma_\mu}_{\gamma \dot 
  \beta} \, \bar W^{\dot \beta } {\cal D}_{\gamma} W_\alpha +\cc 
\eeqa

If we introduce a U(1) under which $S$ has charge $+1$, with all
MSSM fields neutral, only
operators~(\ref{op8}-\ref{op12},~\ref{op14}-\ref{op16},~\ref{op27}-\ref{op30})
are allowed.
Taking the $R$-charge of $S$ to be zero, only
operators~(\ref{op8},~\ref{op9},~\ref{op11},~\ref{op14},~\ref{op28},~\ref{op29})
preserve both the $U(1)$ and the $R$-symmetry.
Note that scalar masses are generated by~(\ref{op28}),
gaugino masses are generated by~(\ref{op27}) and $A$-terms are
generated by~(\ref{op7}).


\end{document}